# Generalized Linear Weights for Sharing Credits Among Multiple Authors

Ash Mohammad Abbas
Department of Computer Engineering
Aligarh Muslim University
Aligarh - 202002, India
Email: am.abbas.ce@amu.ac.in

*Abstract*—Assignment of weights to multiple authors of a paper is a challenging task due to its dependence on the conventions that may be different among different fields of research and research groups. In this paper, we describe a scheme for assignment of weights to multiple authors of a paper. In our scheme, weights are assigned in a linearly decreasing/increasing fashion depending upon the weight decrement/increment parameter. We call our scheme *Arithmetic: Type-2* scheme as the weights follow an arithmetic series. We analyze the proposed weight assignment scheme and compare it with the existing schemes such as equal, arithmetic, geometric, and harmonic. We argue that the a positional weight assignment scheme, called arithmetic scheme, which we refer to *Arithmetic: Type-1* in this paper, and the equal weight assignment scheme can be treated as special cases of the proposed *Arithmetic: Type-2* scheme.

## I. INTRODUCTION

Sometimes, one needs to determine the quality of research produced by an author, specifically, for allocating grants or evaluation for selection and/or promotion of a researcher, and for that purpose one needs to have some mechanism. The citations of papers produced by the author may provide some insights and may enable one to devise some scheme or mechanism for evaluating the quality of research produced by the author. The authors can be ranked or indexed based on the indexing mechanism. Most of the mechanisms for evaluating the quality of research are based on the number of citations of papers authored by a researcher.

However, if a paper is written by multiple authors, then it is not the only one author who should receive the full credit, and all authors of the paper should not be given full credits for a multi-authored paper in comparison to a paper which is written by a single author. In other words, the citations of a multi-authored paper should be shared among all authors of the paper, and the indexing technique should be able to incorporate the effect of multiple authorship.

Ideally, sharing the credits should depend upon contributions of individual authors to the paper. An obvious choice can be that the citations are divided equally among all authors of the paper. However, if some of the authors have put more effort into the paper, then it may seem unfair to them. Similarly, if all authors contributed to a paper almost equally, then it would be unfair to divide the credits unequally. We would like to point out that there is no universally agreed upon convention by which one can determine the extent of contributions of individual authors of the paper as the conventions may be different for different areas of research and research groups.

Assume that there is a research area where all authors do not contribute equally, and authors follow a convention that there names in the papers produced by them appear in decreasing order of their contributions. (If it seems hypothetical, assume that within a research area, there is a research team that adopts this convention.). Though this is an unsupported assumption, one may find examples of such teams in real life. For such a hypothetical research team, there should be a scheme that assigns weights to multiple authors of the paper in decreasing order of their positions. A scheme that assigns the weights to authors according to their positions is described in [2]. In [3], a scheme for sharing the credits harmonically is emphasized over equal, geometric, and arithmetic sharings.

In this paper, we propose a scheme for assigning weights for multiple authors of a paper. We call it *Arithmetic: Type-2* weight assignment scheme as the weights follow an arithmetic series. We refer to the positional weight assignment scheme described in [1] as *Arithmetic: Type-1* scheme. Our scheme assigns the weight to authors in linearly decreasing order and is capable of assigning the weights in linearly increasing order as well depending upon the weight decrement/increment parameter. The scheme is generalized in the sense that equal weight assignment scheme and the *Arithmetic: Type-1* weight assignment scheme can be considered as special cases of the proposed scheme.

The rest of this paper is organized as follows. In section II, we describe the proposed scheme. In section III, we compare the proposed scheme with other weight assignment schemes. The last section is for conclusion and future work.

## II. GENERALIZED LINEAR WEIGHTS

In this section, we describe the proposed weight assignment scheme. The definition of the proposed weights is as follows.

*Definition 1:* Let the number of authors of a paper be $k$. Let $w_1$ and $w_k$ be the weights of the first and last authors of the paper. Let there be a weight assignment policy, say $A$, which assigns the following weight to the $j$th author.

$$w_j = w_1 - (j-1)\alpha \qquad (1)$$

TABLE I
POSITION OF AUTHORS AND CORRESPONDING WEIGHTS IN THE PROPOSED *Arithmetic: Type-2* SCHEME.

| Position | Weight | | |
|---|---|---|---|
| 1 | $w_1$ | | |
| 2 | $w_2$ | = | $w_1 - \alpha$ |
| 3 | $w_3$ | = | $w_2 - \alpha$ |
|  |  | = | $w_1 - 2\alpha$ |
| $\vdots$ | $\vdots$ | | |
| $j$ | $w_j$ | = | $w_{j-1} - \alpha$ |
|  |  | = | $w_1 - (j-1)\alpha$ |
| $\vdots$ | $\vdots$ | | |
| $k-1$ | $w_{k-1}$ | = | $w_{k-2} - \alpha$ |
|  |  | = | $w_1 - (k-2)\alpha$ |
| $k$ | $w_k$ | = | $w_{k-1} - \alpha$ |
|  |  | = | $w_1 - (k-1)\alpha$ |

where, $2 \leq j \leq k$, $0 \leq \alpha \leq 1$, and $\sum_{j=1}^{k} w_j = 1$. In other words, the weights assigned to authors 1 to $k$ are given in Table I.

We call $\alpha$, $0 \leq \alpha \leq 1$, the *weight decrement parameter* because of its use in decrementing the weights from the first author to the last author.

Note that (1) represents an arithmetic series, therefore, we call these weights as *Arithmetic: Type-2* weights. These weights are named *Arithmetic: Type-2* weights because there is another weight assignment scheme described in [1] and the weights, therein, also form an arithmetic series. To differentiate between the two types of weights, we call the weights described in [1] as *Arithmetic: Type-1* and the weights proposed in this paper as *Arithmetic: Type-2* weights. Later in this paper, we shall show that *Type-2* weights can be considered as an special case of *Type-1* weights. Based on the weights defined above, we prove a lemma that provides the weights of the first and last authors in terms of the number of authors, $k$, and the weight decrement parameter, $\alpha$.

*Lemma 1:* Let there be $k$ number of author(s) of a paper, and the weight $j$th author be defined as $w_j = w_{j-1} - \alpha$, where $0 \leq \alpha \leq 1$. Let $w_1$ and $w_k$ be the weights of the first and the last authors, which can be expressed as follows.

$$w_1 = \frac{1}{k} + \frac{\alpha(k-1)}{2}, \qquad (2)$$

and,

$$w_k = \frac{1}{k} - \frac{\alpha(k-1)}{2}. \qquad (3)$$

*Proof:* For any weight assignment policy, we have,

$$\sum_{j=1}^{k} w_j = 1. \qquad (4)$$

Using the values of $w_j$, we have,

$$\sum_{j=1}^{k} w_1 - (j-1)\alpha = 1. \qquad (5)$$

Or,

$$kw_1 - \alpha \sum_{j=1}^{k-1} j = 1. \qquad (6)$$

Or,

$$kw_1 - \alpha \left\{ \frac{k(k-1)}{2} \right\} = 1. \qquad (7)$$

Solving it for $w_1$, we have,

$$w_1 = \frac{1}{k} + \frac{\alpha(k-1)}{2}.$$

This is same as (2). To prove (3), we know that the weight of $k$th author is given by

$$w_k = w_1 - (k-1)\alpha. \qquad (8)$$

Using (2) and (8), we have,

$$w_k = \frac{1}{k} - \frac{\alpha(k-1)}{2}.$$

This is the same as (3), and this completes the proof. ∎

Lemma 1 relates the weights of the first author and the last author with the number of authors of the paper, $k$, and the weight decrement parameter, $\alpha$. Based on the Lemma 1, we have the following corollary.

*Corollary 1:* The ratio of the weights of the first author and the last author under the proposed weight assignment scheme is given by

$$\frac{w_1}{w_k} = \frac{\frac{1}{k} + \frac{\alpha(k-1)}{2}}{\frac{1}{k} - \frac{\alpha(k-1)}{2}} \qquad (9)$$

*Proof:* The proof of Corollary 1 is straight forward by simply using (2) and (3). ∎

In other words, the value of $\frac{w_1}{w_k}$ depends upon the number of authors, $k$, and the weight decrement parameter, $\alpha$.

In general, the weight of $j$th author can be expressed in terms of $k$ and $\alpha$ as described in the following lemma.

*Lemma 2:* Let the number of author(s) of a paper be $k$, and the weight decrement parameter be $\alpha$ such that $0 \leq \alpha \leq 1$. Let the weight of $j$th author of the paper be defined as $w_j = w_{j-1} - \alpha$. Then, the weight of $j$th author can be expressed as follows.

$$w_j = \frac{1}{k} + \frac{k - 2j + 1}{2}\alpha. \qquad (10)$$

*Proof:* For $j$th author, $1 \leq j \leq k$, we have,

$$w_j = w_1 - (j-1)\alpha. \qquad (11)$$

Using (2) and (11), we have,

$$w_j = \frac{1}{k} + \frac{\alpha(k-1)}{2} - (j-1)\alpha. \qquad (12)$$

Simplifying the R.H.S. of (12), we get,

$$w_j = \frac{1}{k} + \frac{k - 2j + 1}{2}\alpha.$$

∎

In what follows, we consider examples to show how the weights are assigned.

*Example 1:* Let there be two authors who are assigned weights $w_1$ and $w_2$. Each author can be assigned an equal weight of 0.5, the parameter $\alpha = 0$. In an unequal weight assignment scheme, the weights can be assigned from the following weight sets $\{0.52, 0.48\}$, $\{0.55, 0.45\}$, $\{0.60, 0.40\}$, $\{0.65, 0.35\}$, $\{0.70, 0.30\}$, and so on, depending upon how much weights one wishes to assign to the first author and to the second author. The values of $\alpha$ are $0.04, 0.1, 0.2, 0.3$, and $0.4$, respectively. One is not restricted to only these weight sets, and can form other weight sets as well. If $w_1 = w$, then $w_2 = 1 - w$. If one wishes to assign a larger weight to the second author, one can simply reverse the order of the elements of the weight sets.

We now consider another example to show how the weights are assigned to multiple authors.

*Example 2:* Let there be three authors with weights $w_1$, $w_2$, $w_3$. A possible assignment can be $\{0.433, 0.333, 0.233\}$, where the vale of $\alpha$ is $0.1$. Another assignment can be $\{0.383, 0.333, 0.283\}$ using the value of $\alpha$ to be equal to $0.05$. Besides these, there can be other assignments as well depending upon the value of $\alpha$.

We would like to mention that these weight assignments are carried out using (10) for a given value of the number of authors, $k$, and the weight decrement parameter, $\alpha$.

Given the values of $k$, $w_1$ and $w_k$, one may compute the value of weight decrement parameter, $\alpha$. For that purpose, one needs to express $\alpha$ as a function of these parameters. Using (8), we can write,

$$\alpha = \frac{w_1 - w_k}{k - 1}. \quad (13)$$

For $j$th author, where $1 \leq j \leq k$, we have from (11),

$$w_j = w_1 - (j - 1)\alpha.$$

Using (13) and (11), we have,

$$w_j = w_1 - (j - 1)\left\{\frac{w_1 - w_k}{k - 1}\right\}. \quad (14)$$

The equation (14) can be simplified to yield

$$w_j = \frac{(k - j)w_1 + (j - 1)w_k}{k - 1}. \quad (15)$$

We now prove a lemma that relates the threshold on the weight of the last author and the decrement parameter, $\alpha$.

*Lemma 3:* Let the minimum value allowed for $w_k$ be $\mu$, then $\alpha$ should be chosen in such a fashion so that the following inequality is satisfied.

$$\alpha \leq \frac{2}{(k - 1)}\left\{\frac{1}{k} - \mu\right\}. \quad (16)$$

*Proof:* Given that $w_k \geq \mu$. Using (8), we have,

$$\frac{1}{k} - \frac{\alpha(k - 1)}{2} \geq \mu. \quad (17)$$

Rewriting (17), we have,

$$\frac{k - 1}{2}\alpha \leq \left(\frac{1}{k} - \mu\right). \quad (18)$$

Therefore, we get,

$$\alpha \leq \frac{2}{k - 1}\left(\frac{1}{k} - \mu\right).$$

∎

Based on Lemma 3, we have the following corollary.

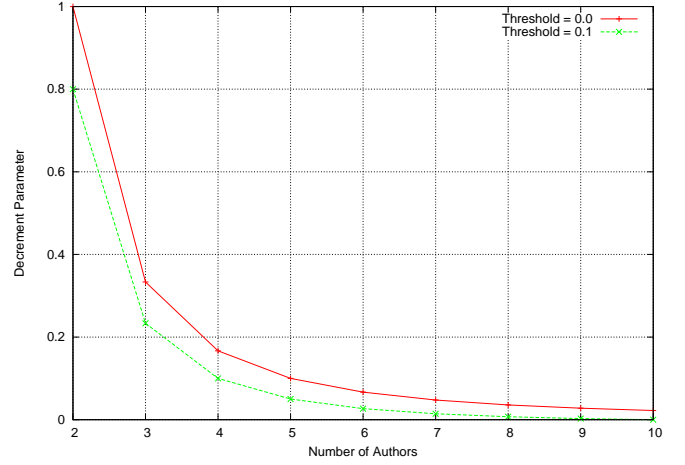

Fig. 1. The decrement parameter, $\alpha$, as a function of the number of authors for different values of the threshold, $\mu$, on the weight of the last author.

*Corollary 2:* The weights should be assigned in such a fashion so that $w_k > 0$. This holds when the following condition is satisfied.

$$\alpha < \frac{2}{k(k - 1)}, \quad k \geq 2. \quad (19)$$

*Proof:* The proof of Corollary 2 is straight forward. One can simply put $\mu = 0$ in (17) to obtain (19). ∎

Figure 1 shows the decrement parameter, $\alpha$, as a function of the number of authors for different values of the threshold, $\mu$, on the weight of the last author, $w_k$.

## III. COMPARISON WITH OTHER WEIGHT ASSIGNMENT SCHEMES

In this section, we compare the proposed weight assignment scheme with other weight assignment schemes.

### A. Comparison with Equal Weights

Note that depending the values of the decrement parameter, $\alpha$, we can vary the weights of the authors. For $\alpha = 0$, all authors are assigned equal weights. The following lemma gives the relationship between linear weights described in this paper and the equal weight assigned to all authors of the paper.

*Lemma 4:* Let there be $k$ number of authors of a paper and the weight decrement parameter be $\alpha$, $0 \leq \alpha \leq 1$. The amount of increase/decrease in the weight of $j$th author of the paper under linear weight assignment scheme as compared to the equal weight assignment scheme is given by the following expression.

$$\Delta_j = \frac{k - 2j + 1}{2}\alpha. \quad (20)$$

*Proof:* The proof of the Lemma 4 is straight forward using (10), where $\frac{1}{k}$ in R.H.S. represents the weight of each author under equal weight assignment scheme; rearranging it gives (20). ∎

We would like to point out that equal weights are also linear weights, and can be represented by a horizontal line. The distance between the $x$-axis (corresponding to author positions for a paper) and the horizontal line is $\frac{1}{k}$, and the slope is $0$.

TABLE II
NUMBER OF AUTHORS AND *Arithmetic: Type-1* WEIGHTS OF INDIVIDUAL AUTHORS.

| Number of Authors | $w_1$ | $w_2$ | $w_3$ | $w_4$ | $w_5$ | $w_6$ | $w_7$ | $w_8$ | $w_9$ | $w_{10}$ |
|---|---|---|---|---|---|---|---|---|---|---|
| 1 | $1$ | | | | | | | | | |
| 2 | $\frac{2}{3}$ | $\frac{1}{3}$ | | | | | | | | |
| 3 | $\frac{3}{6}$ | $\frac{2}{6}$ | $\frac{1}{6}$ | | | | | | | |
| 4 | $\frac{4}{10}$ | $\frac{3}{10}$ | $\frac{2}{10}$ | $\frac{1}{10}$ | | | | | | |
| 5 | $\frac{5}{15}$ | $\frac{4}{15}$ | $\frac{3}{15}$ | $\frac{2}{15}$ | $\frac{1}{15}$ | | | | | |
| 6 | $\frac{6}{21}$ | $\frac{5}{21}$ | $\frac{4}{21}$ | $\frac{3}{21}$ | $\frac{2}{21}$ | $\frac{1}{21}$ | | | | |
| 7 | $\frac{7}{28}$ | $\frac{6}{28}$ | $\frac{5}{28}$ | $\frac{4}{28}$ | $\frac{3}{28}$ | $\frac{2}{28}$ | $\frac{1}{28}$ | | | |
| 8 | $\frac{8}{36}$ | $\frac{7}{36}$ | $\frac{6}{36}$ | $\frac{5}{36}$ | $\frac{4}{36}$ | $\frac{3}{36}$ | $\frac{2}{36}$ | $\frac{1}{36}$ | | |
| 9 | $\frac{9}{45}$ | $\frac{8}{45}$ | $\frac{7}{45}$ | $\frac{6}{45}$ | $\frac{5}{45}$ | $\frac{4}{45}$ | $\frac{3}{45}$ | $\frac{2}{45}$ | $\frac{1}{45}$ | |
| 10 | $\frac{10}{55}$ | $\frac{9}{55}$ | $\frac{8}{55}$ | $\frac{7}{55}$ | $\frac{6}{55}$ | $\frac{5}{55}$ | $\frac{4}{55}$ | $\frac{3}{55}$ | $\frac{2}{55}$ | $\frac{1}{55}$ |

TABLE III
NUMBER OF AUTHORS AND GEOMETRIC WEIGHTS OF INDIVIDUAL AUTHORS.

| Number of Authors | $w_1$ | $w_2$ | $w_3$ | $w_4$ | $w_5$ | $w_6$ | $w_7$ | $w_8$ | $w_9$ | $w_{10}$ |
|---|---|---|---|---|---|---|---|---|---|---|
| 1 | $1$ | | | | | | | | | |
| 2 | $\frac{2}{3}$ | $\frac{1}{3}$ | | | | | | | | |
| 3 | $\frac{4}{7}$ | $\frac{2}{7}$ | $\frac{1}{7}$ | | | | | | | |
| 4 | $\frac{8}{15}$ | $\frac{4}{15}$ | $\frac{2}{15}$ | $\frac{1}{15}$ | | | | | | |
| 5 | $\frac{16}{31}$ | $\frac{8}{31}$ | $\frac{4}{31}$ | $\frac{2}{31}$ | $\frac{1}{31}$ | | | | | |
| 6 | $\frac{32}{63}$ | $\frac{16}{63}$ | $\frac{8}{63}$ | $\frac{4}{63}$ | $\frac{2}{63}$ | $\frac{1}{21}$ | | | | |
| 7 | $\frac{64}{127}$ | $\frac{32}{127}$ | $\frac{16}{127}$ | $\frac{8}{127}$ | $\frac{4}{127}$ | $\frac{2}{127}$ | $\frac{1}{127}$ | | | |
| 8 | $\frac{128}{255}$ | $\frac{64}{255}$ | $\frac{32}{255}$ | $\frac{16}{255}$ | $\frac{8}{255}$ | $\frac{4}{255}$ | $\frac{2}{255}$ | $\frac{1}{255}$ | | |
| 9 | $\frac{256}{511}$ | $\frac{128}{511}$ | $\frac{64}{511}$ | $\frac{32}{511}$ | $\frac{16}{511}$ | $\frac{8}{511}$ | $\frac{4}{511}$ | $\frac{2}{511}$ | $\frac{1}{511}$ | |
| 10 | $\frac{512}{1023}$ | $\frac{256}{1023}$ | $\frac{128}{1023}$ | $\frac{64}{1023}$ | $\frac{32}{1023}$ | $\frac{16}{1023}$ | $\frac{8}{1023}$ | $\frac{4}{1023}$ | $\frac{2}{1023}$ | $\frac{1}{1023}$ |

### B. Comparison with Arithmetic: Type-1 Weights

A positional weight assignment scheme is described in [2], where the weight of $j$th author is given by the following expression.

$$w_j = \frac{2(k-j+1)}{k(k+1)}. \quad (21)$$

As mentioned earlier, we call these weights as *Arithmetic: Type-1* weights and the weights proposed in this paper as *Arithmetic: Type-2* weights. In the following, we state a lemma that relates the weight of the first author and the last author under *Arithmetic: Type-1* weight assignment scheme.

*Lemma 5:* Let $k$ be the number of authors in a given paper. The weight of the first author under *Arithmetic: Type-1* weight assignment scheme is $k$ times the weight of the last author.

*Proof:* Using (21) the weights of the first and the last authors of the paper are as follows.

$$w_1 = \frac{2}{k+1}, \quad \text{and} \quad w_k = \frac{2}{k(k+1)}. \quad (22)$$

Therefore,

$$\frac{w_1}{w_k} = k. \quad (23)$$

∎

We now prove a lemma to relate the *Arithmetic: Type-1* weights described in this paper to the *Arithmetic: Type-2*

TABLE IV
NUMBER OF AUTHORS AND HARMONIC WEIGHTS OF INDIVIDUAL AUTHORS.

| Number of Authors | $w_1$ | $w_2$ | $w_3$ | $w_4$ | $w_5$ | $w_6$ | $w_7$ | $w_8$ | $w_9$ | $w_{10}$ |
|---|---|---|---|---|---|---|---|---|---|---|
| 1 | 1 | | | | | | | | | |
| 2 | $\frac{2}{3}$ | $\frac{1}{3}$ | | | | | | | | |
| 3 | $\frac{6}{11}$ | $\frac{3}{11}$ | $\frac{2}{11}$ | | | | | | | |
| 4 | $\frac{12}{25}$ | $\frac{6}{25}$ | $\frac{4}{25}$ | $\frac{3}{25}$ | | | | | | |
| 5 | $\frac{60}{137}$ | $\frac{30}{137}$ | $\frac{20}{137}$ | $\frac{15}{137}$ | $\frac{12}{137}$ | | | | | |
| 6 | $\frac{60}{147}$ | $\frac{30}{147}$ | $\frac{20}{147}$ | $\frac{15}{147}$ | $\frac{12}{147}$ | $\frac{10}{147}$ | | | | |
| 7 | $\frac{420}{1089}$ | $\frac{210}{1089}$ | $\frac{140}{1089}$ | $\frac{105}{1089}$ | $\frac{84}{1089}$ | $\frac{70}{1089}$ | $\frac{60}{1089}$ | | | |
| 8 | $\frac{840}{2283}$ | $\frac{420}{2283}$ | $\frac{280}{2283}$ | $\frac{210}{2283}$ | $\frac{168}{2283}$ | $\frac{140}{2283}$ | $\frac{120}{2283}$ | $\frac{105}{2283}$ | | |
| 9 | $\frac{2520}{7129}$ | $\frac{1260}{7129}$ | $\frac{840}{7129}$ | $\frac{630}{7129}$ | $\frac{504}{7129}$ | $\frac{420}{7129}$ | $\frac{360}{7129}$ | $\frac{315}{7129}$ | $\frac{280}{7129}$ | |
| 10 | $\frac{2520}{7379}$ | $\frac{1260}{7379}$ | $\frac{840}{7379}$ | $\frac{630}{7379}$ | $\frac{504}{7379}$ | $\frac{420}{7379}$ | $\frac{360}{7379}$ | $\frac{315}{7379}$ | $\frac{280}{7379}$ | $\frac{252}{7379}$ |

weights described in [2].

*Lemma 6:* Let there be $k$ number of authors of a paper. Let there be a linear weight assignment scheme named *Arithmetic: Type-2* with weight decrement parameter $\alpha$. The weight of $j$th author of the paper, where $1 \leq j \leq 1$, under both the *Arithmetic: Type-2* weight assignment and the *Arithmetic: Type-1* weight assignment schemes comes out to be the same for the following value of $\alpha$.

$$\alpha = \frac{2}{k(k+1)}. \tag{24}$$

*Proof:* Using (10) and (21), we have,

$$\frac{1}{k} + \frac{k-2j+1}{2}\alpha = \frac{2(k-j+1)}{k(k+1)}. \tag{25}$$

Or,

$$\frac{k-2j+1}{2}\alpha = \frac{2(k-j+1)}{k(k+1)} - \frac{1}{k}. \tag{26}$$

Or,

$$\frac{k-2j+1}{2}\alpha = \frac{k-2j+1}{k(k+1)}. \tag{27}$$

Therefore,

$$\alpha = \frac{2}{k(k+1)}$$

which is same as (24). This completes the proof. ■

We pointed out that equal weights are linear. Let us examine the *Arithmetic: Type-1* weights given by (21). To show that the *Arithmetic: Type-1* weights are also linear, one can rewrite (21) as follows.

$$w_j = \frac{-2j}{k(k+1)} + \frac{2}{k}. \tag{28}$$

Comparing (28) with $y = mx + c$, we have $m = -\frac{2}{k(k+1)}$ and $c = \frac{2}{k}$. Therefore, *Arithmetic: Type-1* weights given by (21) are also linear. Further, the proposed *Arithmetic: Type-2* weights are also positional weights. Since both *Type-1* and *Type-2* are linear and positional weights, one may ask a question about the difference between the two schemes. To answer it, we would like to mention that the proposed *Arithmetic: Type-2* weight assignment is generalized in the sense that the *Arithmetic: Type-1* weight assignment scheme as given by (21) is a special case of the proposed *Arithmetic: Type-2* scheme. In case of *Type-2* weights, one can vary the weights assigned from first through the last author by simply varying the weight decrement parameter $\alpha$, however, no such control is there for *Type-1* weight as given by (21). In case of *Type-1* weights as given by (21), the weights are fixed from the first through the last author for a given number of authors of the paper and cannot be varied. On the other hand, comparing the proposed *Type-2* scheme with the equal weight assignment scheme, we pointed out earlier that equal weight assignment scheme can also be considered as a special case of the proposed weight assignment scheme. Further by taking a negative value of $\alpha$, one can have linearly increasing *Type-2* weights in case the convention followed in some research discipline is that authors shall be assigned linearly increasing weights. Table II shows the weights of individual authors in the *Arithmetic: Type-1* weight assignment scheme.

### C. Comparison with Geometric Weights

Let $k$ be the number of authors of a paper, the geometric weight for $j$th author is defined as follows.

$$w_j = \frac{2^{k-j}}{2^k - 1}. \tag{29}$$

Note that geometric weights for a given number of authors are positional and nonlinear weights; and the weight of the last author of the paper decreases exponentially with the number of authors. The following lemma provides the ratio of the weights of the first author and the last author.

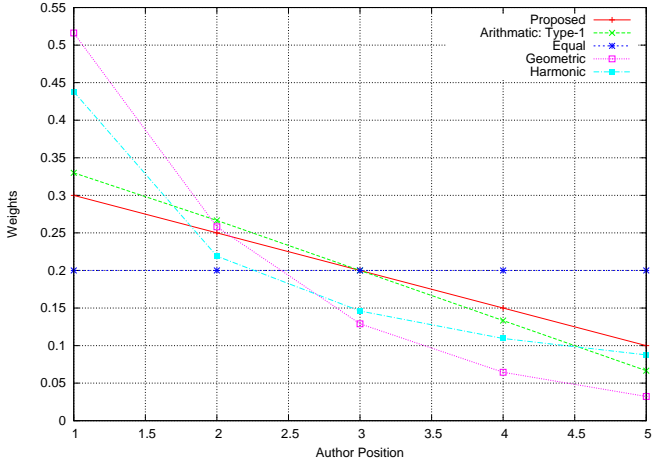

Fig. 2. Weights of authors as a function of author positions under different weight assignment scheme for 5 authors. The weights of authors under the proposed (Arithmetic: Type 2) scheme are corresponding to the value of $\alpha = 0.05$.

TABLE V
FEATURES OF DIFFERENT WEIGHT ASSIGNMENT SCHEMES.

| Scheme | $\frac{w_1}{w_k}$ | Feature-1 | Feature-2 | Weights |
|---|---|---|---|---|
| Equal | 1 | Linear | Position independent | Fixed |
| Geometric | $2^{k-1}$ | Linear | Positional | Fixed |
| Harmonic | $k$ | Non-linear | Positional | Fixed |
| Arithmetic: Type-1 | $k$ | Non-linear | Positional | Fixed |
| Arithmetic: Type-2 (proposed weights) | Variable | Linear | Positional | Variable |

*Lemma 7:* Let there be $k$ number of authors of a paper. The weight of the first author is $2^{k-1}$ times the weight of the last author under geometric weight assignment scheme.

*Proof:* Using (29), the weights of the first author and the last author are given by the following expressions.

$$w_1 = \frac{2^{k-1}}{2^k - 1}, \text{ and } w_k = \frac{1}{2^k - 1}. \quad (30)$$

Therefore,

$$\frac{w_1}{w_k} = 2^{k-1}. \quad (31)$$

For example, if there are 4 authors, the weight of the first author is 8 times the weight of the last author. On the other hand, in case of proposed weight assignment scheme the ratio of the wights of the first author and the last author is given by (9), which depends on the number of authors of the paper, $k$, and the weight decrement parameter. For $k = 4$, and $\alpha = 0.1$, the weights of authors are $\{0.4, 0.3, 0.2, 0.1\}$, i.e. the weight of the first author is only 4 times the weight of the last author. In other words, for the given value of $\alpha$, the weight of the last author in case of proposed scheme is twice of the weight assigned in geometric weight assignment scheme. Table III shows the geometric weights for different number of authors.

### D. Comparison with Harmonic Weights

Let there be $k$ number of authors, the weight of $j$th author under harmonic weight assignment scheme is as follows.

$$w_j = \frac{\frac{1}{j}}{\sum_{i=1}^{k} \frac{1}{i}}. \quad (32)$$

Note that the denominator of the harmonic weights is the harmonic series which can be written as follows.

$$\sum_{i=1}^{k} \frac{1}{i} = \ln k + \gamma + \epsilon_k. \quad (33)$$

where, $\epsilon_k \sim \frac{1}{2k}$, and $\gamma \approx 0.5772$ is called Euler-Mascheroni constant. Also, we would like to point out that $\ln(k+1) < \sum_{i=1}^{k} \frac{1}{i} < \ln k + 1$. Using (32) and (33), we get,

$$w_j = \frac{\frac{1}{j}}{\ln k + \gamma + \epsilon_k}. \quad (34)$$

We now state a lemma that relates the weights under harmonic and equal weight assignment scheme.

*Lemma 8:* The amount of increase/decrease in the weight of $j$th author under harmonic weight assignment scheme as compared to the equal weight assignment scheme is

$$\delta_j = \frac{k - j(\ln k + \gamma + \epsilon_k)}{jk(\ln k + \gamma + \epsilon_k)}. \quad (35)$$

*Proof:* Using (34) and the expression for equal weights among $k$ authors, we have,

$$\delta_j = \frac{\frac{1}{j}}{\ln k + \gamma + \epsilon_k} - \frac{1}{k}. \quad (36)$$

Simplifying it, we get (35). ∎

The harmonic weights for individual authors are shown in Table III. We state a lemma that relates the weight of the first author and the last author under geometric weight assignment scheme, which is as follows.

*Lemma 9:* Let $k$ be the number of authors of a paper under harmonic weight assignment scheme. The weight of the first author is $k$ times the weight of the last author under harmonic weight assignment scheme.

*Proof:* Using (32), the weights of the first and last authors under the harmonic weight assignment scheme are as follows.

$$w_1 = \frac{1}{\sum_{i=1}^{k} \frac{1}{i}}, \text{ and } w_k = \frac{\frac{1}{k}}{\sum_{i=1}^{k} \frac{1}{i}}. \quad (37)$$

Therefore,

$$\frac{w_1}{w_k} = k. \quad (38)$$

∎

### E. Summary of Comparisons

Note that the proposed weights together with geometric and harmonic weights are positional weights as they can be used to assign weights to authors of a paper based on their ranks or positions. The equal weights are not positional as the weight assigned to all authors is $\frac{1}{k}$ irrespective of the position of authors. Further, the proposed weights and equal weights are linear weights as the weights assigned to authors of the same paper lie on a line. However, the geometric and the harmonic weights are nonlinear weights as the weights assigned to authors of the same paper may not lie on a line. To illustrate it, weights under different weight assignment schemes are shown in Figure 2 for $k = 5$, where the value of $\alpha$ is $0.05$ for the proposed weight assignment scheme, which we call *Arithmetic: Type-2*. Note that the slope of the proposed *Arithmetic: Type-2* weights can be made more gradual by choosing a relatively small value of $\alpha$. However, there is no such control over other types of weights. Table V summarizes the features of different weight assignment schemes.

As far as the applicability of the weights is concerned, we suggest that the citations of a multi-authored paper should be multiplied by the corresponding weights of the individual authors and then the index of individual authors can be computed following the normal procedure of computing the corresponding index. This can be applied for all types of weights.

## IV. Conclusion

Devising a scheme for assigning weights to multiple authors of a paper is a challenging task due to the absence of a universally agreeable set of conventions. The conventions may differ among the research fields and even among different research teams within an area of research. In this paper, we proposed a scheme for assigning weights to multiple authors of a paper. Our scheme assigns the weights to authors in a linearly decreasing/increasing order depending the value of the weight decrement/increment parameter. We refer to the proposed scheme as *Arithmetic: Type-2* and the scheme described in [1] as *Arithmetic: Type-1* scheme, as both the schemes follow an arithmetic progression. The proposed *Arithmetic: Type-2* weight assignment scheme is generalized in the sense that equal weight assignment scheme and the *Arithmetic: Type-1* weight assignment scheme can be treated as special cases of the proposed *Type-2* scheme. Further, we compared the proposed scheme with other existing weight assignment schemes such as geometric and harmonic schemes. We observed that our scheme is flexible in assigning the weights as the weights can be varied by varying the weight decrement/increment parameter, however, no such control exists for other schemes. Our scheme can be incorporated into an index by simply multiplying the citations of a multi-authored paper with the weights of the corresponding authors and then the index of individual authors can be computed. Incorporating them into an existing index, proposal of a new or modified index, and the validations in a variety of fields form the future works.